\documentclass[prb,twocolumn,showpacs,floatfix,aps]{revtex4}
\usepackage{graphicx}
\usepackage{epsfig}
\newcommand{\etal}{{\it et~al.}}

\begin{document}

\title{Excess current in point contacts on two-band superconductor MgB$_2$ in magnetic field}
\author{Yu. G. Naidyuk,\footnote{e-mail: naidyuk@ilt.kharkov.ua} O. E. Kvitnitskaya,
I. K. Yanson}

\affiliation {B.Verkin Institute for Low Temperature Physics and
Engineering, National Academy  of Sciences of Ukraine, 47 Lenin
Ave., 61103, Kharkiv, Ukraine}

\author{S. Lee, and S. Tajima}
\affiliation{Superconductivity Research Laboratory, ISTEC, 1-10-13
Shinonome, Koto-ku, Tokyo 135-0062, Japan}

\date{\today}
\begin{abstract}
A series of $I(V)$ characteristics and bias-dependent differential
resistances $dV/dI(V)$ of the point contacts made of the single
crystal of two-band superconductor MgB$_2$ were measured in
magnetic fields up to 9\,T. The magnetic field dependences of the
excess current in the $I(V)$ curves were obtained and analyzed
using Koshelev and Golubov's [Phys. Rev. Lett. {\bf 90}, 177002
(2003)] theoretical results for the mixed state of a dirty
two-band superconductor. Introducing a simple model for the excess
current in the point contact in the mixed state, our data can be
qualitatively described using the theoretical magnetic field
dependence of the superconducting order parameter of the $\sigma$
and $\pi$-bands and the averaged electronic density of states in
MgB$_2$.

\pacs{74.45.+c, 74.70Ad, 73.40.Jn}
\end{abstract}

\maketitle

\section{Introduction}

MgB$_2$ has attracted considerable attention due to its high T$_c$
($\approx$40\,K) which is due to specific multi-band electronic
structure (see Ref.\,\cite{Mazin} and Refs. therein). Here
superconductivity develops in two-dimensional cylindrical sheets
of the Fermi surface constituted from the $\sigma$-band and
three-dimensional tubular Fermi surface from the $\pi$-band. The
commonly accepted explanation for high-T$_c$  superconductivity in
MgB$_2$ is connected with the strong interaction between the
charge carriers and the $E_{2g}$ phonon mode, \cite{Ann} caused by
the antiparallel vibrations of the atoms in the honeycomb-like
boron planes. The superconducting order parameter is distributed
over the Fermi surface of MgB$_2$, being $\Delta_\sigma\approx$
7\,meV for the $\sigma$-band and $\Delta_\pi\approx$ 2\,meV for
the $\pi$-band. \cite{Choi} The different properties of the
$\sigma$ and $\pi$ bands with different anisotropies result in
very peculiar and interesting physical characteristics of MgB$_2$
both in the normal and especially in the superconducting state.
\cite{Mazin,Choi,Buzea}

Point-contact (PC) spectroscopy is one of the straightforward
methods which has proved the existence the double-gap
superconductivity in MgB$_2$ (see, e.\,g.,
Refs.\,\cite{Szabo,Gonnelli,Bobrov,Naidyuk1}). Although much was
done to understand the remarkable superconducting state in this
diboride, some issues  should be clarified, such as the behavior
of the superconducting gaps in the magnetic field and the rapid
depression of the superconducting features of the $\pi$-band in
the PC conductivity by a moderate field of about 1~T. In this
study we carried out PC measurements both of the order parameter
and the excess current versus the magnetic field in MgB$_2$ and
analyzed them using the theory \cite{Koshelev} for the mixed state
of a dirty two-band superconductor.

\section{Samples and measurements}

Our experiments were carried out on the MgB$_2$ single-crystal
samples investigated earlier in Ref.\,\cite{Naidyuk3}. Their
properties are described elsewhere. \cite{Lee} The crystals were
small thin plates (flakes) of a sub-millimeter size. They were
glued with silver epoxy to the sample holder at one of their side
faces. The noble metal counter electrode was brought into a gentle
touch with the opposite side face of the crystal in liquid helium.
Thus we tried to make a contact mainly along the {\it ab} plane,
for which both the $\sigma$ and $\pi$-band superconducting
features should be present in the PC characteristics. The magnetic
field was applied along the c-axis.

The $I(V)$ characteristic and the differential resistance
$dV/dI(V)$ were recorded through sweeping DC current $I$ on which
small AC current $i$ was superimposed and through measuring the
alternating voltage $V_1\propto dV/dI(V)$ using the standard
lock-in technique.

\section{Experimental results}

First we evaluated the characteristic length scales important for
this study. The contact size for the typical resistance of
10\,$\Omega$ is estimated by the Sharvin formula
$d\simeq\sqrt{\rho l/R}$ ($d\simeq 7$\,nm in
Ref.\,\cite{Yansonrev}) irrespectively of the crystallographic
directions. For a dirty constriction (with a very short mean free
path) the use of the Maxwell formula $d\sim\rho/R$ results in $d$
about 0.7\,nm and 2.6\,nm (see Ref.\,\cite{Yansonrev}) for the
$ab$ plane and $c$ direction, respectively. Thus, the contact size
can be of the order of magnitude or smaller than the elastic
electron mean free path ($l_{ab}=24$\,nm, $l_c=6$\,nm, according
to Ref.\,\cite{Eltsev}), which means that theoretically we can
approach the ballistic (spectroscopic) regime. In practice the
spectroscopic regime is confirmed by observation of the
characteristic Andreev-reflection features in $dV/dI$ in the
superconducting state.

The measurements of the superconducting energy gap in MgB$_2$ by
means of the Andreev reflection demonstrate two sets of the
energy-gap minima in the differential resistance d$V/$d$I$ at $\pm
2.4\pm $ 0.1 and $\pm 7.1\pm $0.4\,meV, respectively; their
distribution \cite{Naidyuk1} corresponds to the theoretical
prediction. \cite{Choi}  The d$V/$d$I$ curves can be well fitted
by the known BTK \cite{BTK} equations (with the suitable $\Gamma$
parameter) for two conducting channels with the appropriate weight
factors $w$. \cite{Szabo,Gonnelli,Bobrov,Naidyuk1} The
contribution of the large gap to the double-gap spectra $w_L$ is
the largest along the $ab$ plane and amounts theoretically to
0.34, while it is negligible in the perpendicular direction (along
the $c$ axis). \cite{Brinkman}

\begin{figure}
\includegraphics[width=\columnwidth,angle=0]{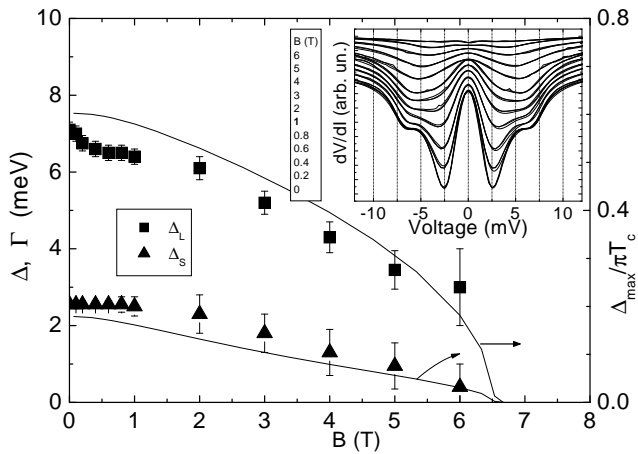}
\caption[]{Magnetic field dependence of large and small
superconducting energy gaps at 4.2\,K (solid squares and
triangles, left y-axis) obtained by a BTK fitting of the curves
shown in the inset. Solid curves show magnetic field dependence of
the maximal pair potential (right y-axis) for MgB$_2$ for
$D_1=D_2$ (see Fig.\,\ref{koshel}). Inset: symmetrized
experimental $dV/dI$ curves in magnetic field (solid lines) for
the single crystal MgB$_2$--Cu 2.2\,$\Omega$ junction along the
$ab$ plane with BTK fit (thin lines) from Ref.\,\cite{Yansonrev}.
Two separate sets of the gap minima around $\pm$2.5 and $\pm$7\,
mV are clearly seen in the low fields.} \label{m163}
\end{figure}

A series of $dV/dI$ measured in magnetic field is shown in
Fig.\,\ref{m163}\,(inset) with clearly seen two sets of the gap
minima (double-gap features). The main panel in Fig.\,\ref{m163}
displays the magnetic field dependence of the large and small gaps
obtained by the routine BTK fitting of the curves from the inset.
In spite of the rapid decrease in the intensity of the small gap
minima with the field (a factor of two at 1\,T, see the inset in
Fig.\,\ref{m163}), the small gap is not suppressed by low fields
($\lesssim$ 1\,T), and the estimated critical field 6-7\,T (see
Fig.\ref{m163}) is much higher than that stated \footnote{In
Ref.\,\cite{Gonnelli1}, Gonnelli \etal ~have corrected their
previous claims and they  mentioned that identification of the
magnetic field (at which the $\pi$-band features in $dV/dI$
disappear visually) with the critical field for the $\pi$  band
might not be correct.} in Refs.\,\cite{Gonnelli,Samuely}. Moreover
this field corresponds to $B_{c2}$ determined from the onset
temperature in the broadened resistive transition in single
crystals. \cite{Masui} Similar behavior of the small gap
persisting up to 5\,T was found in Ref.\,\cite{Bugoslavsky}, where
visual disappearance of the $\pi$-gap features in the PC
characteristics was attributed to broadening of the  spectra in
the magnetic field.

\begin{figure}
\includegraphics[width=\columnwidth,angle=0]{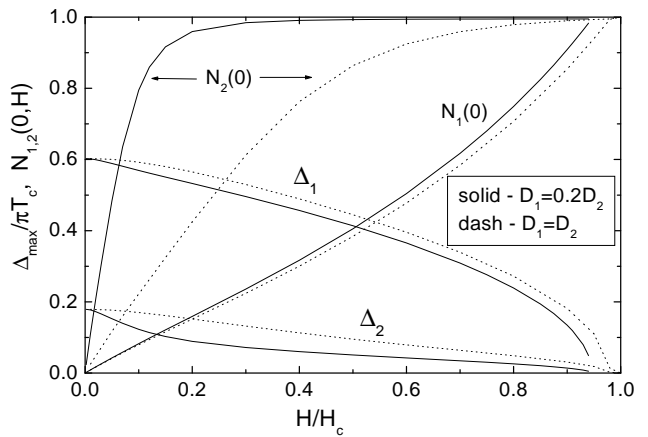}
\caption[]{Magnetic field dependences of the maximal pair
potential $\Delta$ and averaged electron DOS $N(0,H)$ at zero
energy for two ratios of the diffusion constants $D_i=2\pi T_c
\xi_i^2$ for the $\sigma$ (1) and $\pi$ (2) bands. By courtesy of
Koshelev and Golubov. \cite{Koshelev}} \label{koshel}
\end{figure}

Experimental data for $\Delta$ in Fig.\,\ref{m163} are in good
agreement with the magnetic field dependence of the maximal pair
potential $\Delta$ for MgB$_2$ according to Ref.\,\cite{Koshelev}
with $D_1=D_2$  relation. Under this condition, the smaller
$\Delta_2$ (see Fig.\,\ref{koshel}, dashed curve) has a negative
curvature in low and near to the critical fields, which is similar
to the experimental $\Delta_S$ (see Fig.\,\ref{m163}).

The Andreev reflection of electrons from the $N-S$ boundary in a
$N-c-S$ point contact ($N$ stands for the normal metal, $c$ is the
constriction, and $S$ is the superconductor) produces an excess
current $I_{exc}$ in the superconducting state. As a result,
$I(V)$ can be written \cite{BTK} at $eV \gg \Delta$ as $I(V)
=V/R_N+I_{\rm exc}$, where
\begin{equation}
I_{exc}\approx \Delta /eR_N \label{exc}
\end{equation}
with $\Delta $ being the superconducting energy gap and  $R_N$
being the PC resistance in the normal state.

We determined the reduced $I_{exc}(B)/I_{exc}(B=0)$ either from
the raw $I(V)$ or mostly by integrating the reduced
$R^{-1}dV/dI(V)$ or $R\,dI/dV(V)$ after subtracting the normal
state background.  The use of the $I(V)$ curves provides less
accurate results because of the resistance instability for some
contacts during the field sweep.

\begin{figure}
\includegraphics[width=\columnwidth,angle=0]{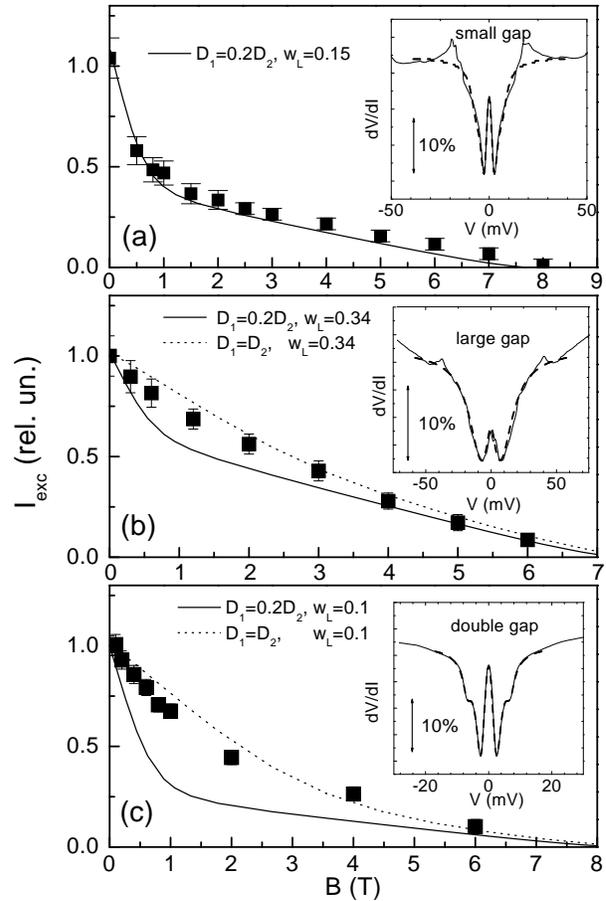}
\caption[]{$I_{exc}$ vs magnetic field behavior for 3 PCs between
the MgB$_2$ single crystal and Cu at 4.2\,K: (a) junction with
small-gap minima and normal state resistance $R$=1.5\,$\Omega$,
(b) junction with large-gap minima from Ref.\,\cite{Naidyuk3},
$R$=6.7\,$\Omega$, (c) junction with double-gap structure from
Fig.\,\ref{m163}, $R$=2.2\,$\Omega$. The lines present calculated
$I_{exc}(B)$ according to (\ref{Ikosh}) with $w_L$, $D_1$ and
$D_2$ shown in each panel. The error bars show difference between
$I_{exc}(B)$ calculated from the experimental (insets, solid
curves) and BTK (insets, dashed curves) $dV/dI$. Insets:
zero-field experimental (solid curves) and BTK (dashed) $dV/dI$
curves for each contact with $\Delta_S$=2.55\,meV and
$\Delta_L$=7.8\,meV (a),  $\Delta_S$=2\,meV and
$\Delta_L$=7.5\,meV (b), and $\Delta_S$=2.55\,meV and
$\Delta_L$=7.1\,meV (c). Vertical bars are drawn to show relative
$dV/dI$ value with respect to the contact resistance $R$. }
\label{IexcB}
\end{figure}

For all investigated contacts the excess current depends on the
magnetic field with the overall positive curvature
(Fig.\,\ref{IexcB}). At first, $I_{exc}(B)$ decreases abruptly and
then more slowly above 1\,T. This corresponds to the
aforementioned drastic suppression of the $dV/dI(V)$ small-gap
minima intensity by a magnetic field about 1\,T and to the
persistence of the residual superconducting minima on a further
increase in the magnetic field. This is quite different from what
is expected for $I_{exc}$ according to (\ref{exc}), taking into
account that $\Delta (B)$ has, as a rule, a negative curvature
(see Fig.\ref{koshel}).

Generally, the excess current, just like  the net current through
the contact, is proportional to the contact area $A$ (see, e.\,g.,
Eqs.\,(18), (19) in Ref.\,\cite{BTK}), i.~e. $I_{exc} \sim
A\times\Delta$. Assuming that $I_{exc}$ is caused by the
superconducting component of the current, which in the mixed state
is determined by the superconducting volume (part) in the PC area,
the coefficient $A$ in the above-mentioned formula should be
effectively reduced by the contribution of the nonsuperconducting
volume. Koshelev and Golubov \cite{Koshelev} calculated the
magnetic field dependence of the electron density of states (DOS)
$N(E)$ at $E$=0 averaged over the vortex unit cell. We assume that
in the first approximation the {\it averaged} $N(0,B)$ can be
regarded as a measure of the nonsuperconducting region in the
mixed state, and, vice versa, $(1-N(0,B))$ characterizes the
superconducting phase portion. Thus, according to above reasoning,
we suppose that
\begin{equation}
I_{exc} \propto w_L(1-N_L(0,B))\Delta_L + w_S(1-N_S(0,B))\Delta_S
 \label{Ikosh}
\end{equation}
with the weight factors $w_L$ and $w_S$ for large and small gaps,
respectively ($w_L+w_S=1$),  $N_{L,S}(0,B)$ and $\Delta(B)_{L,S}$
are taken from Fig.\,\ref{koshel}, where $L\equiv 1$, $S\equiv 2$.
Here, $I_{exc}$ is presented \footnote{According to S. I.
Beloborod`ko`s private communication, $I_{exc}$ is proportional to
$(1-N(0))^{1/2}\Delta$ in the gapless state for a S-c-N junction
with magnetic impurities (see, e.\,g., S.~I.~Beloborod`ko, Low
Temp. Phys., {\bf 29}, 650 (2003) and corresponding Refs.
therein). Therefore, our assumption about the $I_{exc}$ dependence
on the averaged zero-energy density of states has some physical
grounds.} by (\ref{Ikosh}) as a sum of two contribution from the
$\sigma$ and $\pi$ bands weighted by $w_{L,S}$ factors, which
along with $\Delta$ were derived from the fitting of the
Andreev-reflection features in $dV/dI$ at zero field.

Figure\,\ref{IexcB} shows the magnetic field dependence of
$I_{exc}$ for three different types of $dV/dI$ with visually
distinct small, large and double gap structures. Here, we also
plotted $I_{exc}$ calculated by (\ref{Ikosh}). The most unexpected
result is the observation of $dV/dI$ (see Fig.\,\ref{IexcB}(b))
with large-gap minima only, which is discussed in
Ref.\,\cite{Naidyuk3}. The large-gap minima are only slightly
affected by magnetic fields of a few tesla. This is in contrast to
the small-gap minima, which according to
Refs.\,\cite{Szabo,Gonnelli} should vanish above 1\,T. The curve
in the bottom panel was calculated taking $w_L$ from the fitting
of the double-gap structure (see Fig.\,\ref{m163}), while in panel
(a) and (b), where only the small- or large-gap structures are
seen, $w_L$ can vary from zero to its highest value 0.34.
\cite{Brinkman} The latter $w_L$ value was used for the spectra in
the middle panel in which only large-gap minima are seen.
Evaluation of $w_L$ for the spectrum in panel (a) inset is not
straightforward because the spikes present at higher energies mask
the large gap features. However, selected $w_L=0.15$ and
$w_L=0.34$ fit reasonably both zero-field $dV/dI$ and $I_{exc}(B)$
curves in panels (a) and (b), respectively.

In general, the experimental and calculated curves agree well in
spite of the simple empiric model adopted for $I_{exc}$. The
calculated curves describe qualitatively the experimental data,
especially the abrupt initial decrease in $I_{exc}$, pronounced in
the curves with the dominant small-gap features
(Fig.\,\ref{IexcB}(a)). In Fig.\,\ref{IexcB}(c), showing two
curves with different $D_1/D_2$ ratios, it is seen that better
correspondence can be reached with an intermediate ratio between
$D_i$ values. The same occurs for the data in Fig.\,\ref{IexcB}(b)
for the large-gap spectrum, though the experimental data for
large-gap and small-gap spectra are closer to the cases with
$D_1=D_2$. This fact testifies that in this case the $\sigma$ band
is closer to the clean limit, also mentioned in
Ref.\,\cite{Bugoslavsky}.

\section{Conclusions}

We have investigated the excess current in PCs based on single
crystals of MgB$_2$ in the magnetic field and proposed an empiric
model for the excess current behavior in the mixed state. Using
this model, a magnetic field dependence of the excess current was
obtained, which is in qualitative agreement with experimental
data. The anomalous magnetic field dependence of the excess
current in PCs reflects the specific two-band structure of the
superconducting order parameter and the spatial DOS behavior in
MgB$_2$. Thus, $I_{exc}(B)$ can be used for evaluation of the
diffusivity in both bands on the base of Koshelev and Golubov's
theory. \cite{Koshelev}

\section*{Acknowledgements}

We are indebted to A.~E.~Koshelev for providing theoretical data
and L.~V.~Tyutrina for the help in the $dV/dI(V)$ fitting. The
fruitful discussion with S. I. Beloborod`ko, S.-L. Drechsler and
A. N. Omelyanchouk and their comments are highly acknowledged. The
work was partially supported by the National Academy of Sciences
of Ukraine under Project $\Phi$1-19, by the State Foundation of
Fundamental Research of Ukraine under Grant $\Phi$7/528~-~2001 and
by the New Energy and Industrial Technology Development
Organization (NEDO) in Japan. The investigations were carried out
in part using the equipment donated by the Alexander von Humboldt
Foundation (Germany).


\begin{thebibliography}{}

\bibitem{Mazin}  I. I. Mazin, V. P. Antropov, Physica C {\bf 385}, 49
(2003).

\bibitem{Ann} J. M. Ann, W. E. Pickett, Phys. Rev. Lett. {\bf 86},
4366 (2001).

\bibitem{Choi}  Hyoung Joon Choi, David Roundy, Hong Sun, Marvin L. Cohen,
Steven G. Louie, Nature {\bf 418}, 758 (2002).

\bibitem{Buzea} C. Buzea and T. Yamashita, Supercond. Sci. Technol.
{\bf 14}, R115 (2001).

\bibitem{Szabo} P. Szab\'{o}, P.~Samuely, J.~Ka\v{c}mar\v{c}\'{i}k, T.~Klein,
J.~Marcus, D.~Fruchart, S.~Miraglia, C.~Marcenat, A.~G.~M.~Jansen,
Phys. Rev. Lett. {\bf 87} 137005 (2001).

\bibitem{Gonnelli} R. S. Gonnelli, D. Daghero, A. Calzolari, G. A. Ummarino,
V.~A.~Stepanov, J.~Jun, S.~M.~Kazakov and J.~Karpinski,
Phys. Rev. Lett. {\bf 89} 247004 (2002). 

\bibitem{Bobrov} N. L. Bobrov, P. N. Chubov, Yu. G. Naidyuk, L.~V.~Tyutrina,
I.~K.~ Yanson, W.~N.~Kang, Hyeong-Jin Kim, Eun-Mi Choi,
C.~U.~Jung, and Sung-Ik Lee, in book {\it  New Trends in
Superconductivty}, Vol.67 of NATO Science Series II: Math. Phys.
and Chem., ed. by J. F. Annett and S. Kruchinin, (Kluwer Acad.
Publ., 2002), p.225.

\bibitem{Naidyuk1}Yu. G. Naidyuk, I. K. Yanson, L. V. Tyutrina,
N.~L.~Bobrov, P.~N.~Chubov, W.~N.~Kang, Hyeong-Jin Kim, Eun-Mi
Choi, and Sung-Ik Lee, JETP Lett. {\bf 75}, 283 (2002).

\bibitem{Koshelev} A. E. Koshelev, A. A. Golubov, Phys. Rev. Lett. {\bf 90},
177002 (2003).

\bibitem{Naidyuk3}  Yu. G. Naidyuk, I. K. Yanson, O. E. Kvitnitskaya,
S.~Lee, and S.~Tajima, Phys. Rev. Lett., {\bf 90}, 197001 (2003).

\bibitem{Lee} Sergey Lee, Physica C {\bf 385}, 31 41 (2003);
S.~Lee, H.~Mori, T.~Masui, Yu.~Eltsev, A.~Yamanoto and S.~Tajima,
J. Phys. Soc. of Japan, {\bf 70}, 2255 (2001).

\bibitem{Yansonrev} I. K. Yanson, Yu. G. Naidyuk, Fiz. Nizk. Temp.
{\bf 30}, 355 (2004) [Low Temp. Phys. {\bf 30}, 261 (2004)](see
also: cond-mat/0309693).

\bibitem{Eltsev} Yu. Eltsev, S. Lee, K. Nakao, N.~Chikumoto, S.~Tajima,
N.~Koshizuka, and M.~Murakami, Phys. Rev. B, {\bf 65}, 140501(R)
(2002).

\bibitem{BTK} G. E. Blonder, M. Tinkham and T. M. Klapwijk, Phys. Rev. B
{\bf 25}, 4515 (1982).

\bibitem{Brinkman} A. Brinkman, A. A. Golubov, H. Rogalla, O.~V.~Dolgov,
J.~Kortus, Y.~Kong, O.~Jepsen, and O.~K.~Andersen, Phys. Rev. B
{\bf 65}, 180517 (2002).

\bibitem{Samuely} P.~Samuely, P.~Szabo, J.~Kacmarcik, T.~Klein,
A.~G.~M.~Jansen, Physica C {\bf 385} 244 (2003).

\bibitem{Masui} Yu. Eltsev, Physica C {\bf 385}, 162 (2003); T.~Masui,
S. Lee and S. Tajima, Physica C {\bf 383},  299 (2003).

\bibitem{Bugoslavsky} Y.~Bugoslavsky, Y.~Miyoshi, G.~K.~Perkins,
A.~D.~Caplin, L.~F.~Cohen, A.~V.~Pogrebnyakov, X.~X.~Xi, Phys.
Rev. B {\bf 69}, 132508 (2004). 

\bibitem{Gonnelli1} R.~S. Gonnelli, D.~Daghero, G.~A.~Ummarino,
V.~Dellarocca, V.~A.~Stepanov, J.~Jun, S.~M.~Kazakov and J.~Karpinski,
Phys. Rev. B {\bf 69}, 100504(R) (2004). 

\end{thebibliography}
\end{document}